# INTELLIGENT UNIT LEVEL TEST GENERATOR FOR ENHANCED SOFTWARE QUALITY

Ning Luo[1] and Linlin Zhang[2]

[1]Visual Computing Group, Intel Asia-Pacific Research & Development Ltd, Shanghai, China
Ning.luo@intel.com

[2]Visual Computing Group, Intel Asia-Pacific Research & Development Ltd, Shanghai, China
livia.zhang@intel.com

## ABSTRACT

*Unit level test has been widely recognized as an important approach to improve the software quality, as it can expose bugs earlier during the development phase. However, manual unit level test development is often tedious and insufficient. Also, it is hard for developers to precisely identify the most error prone code block deserving the best test coverage by themselves. In this paper, we present the automatic Unit level test framework we used for intel media driver development. It can help us identify the most critical code block, provide the test coverage recommendation, and automatically generate >80% ULT code (~400K Lines of test code) as well as ~35% test cases (~7K test cases) for intel media driver. It helps us to greatly shrink the average ULT development effort from ~24 Man hours to ~3 Man hours per 1000 Lines of driver source code.*

## KEYWORDS

*Unit level test, error prone logic, test coverage inference, automatic ULT generation, fuzzing, condition/decision coverage.*

## 1. INTRODUCTION

Unit level test (ULT) has been widely recognized as an important approach to improving software quality, as it can expose bugs earlier during the development phase. For Intel Media driver, our ULT coverage target is 100% for functional coverage and >70% for conditional coverage.

However, the large scale of Intel Media driver (~1.7 millions of lines of code), as well as the tremendous amount of the classes, methods and possible inputs, makes manual ULT development till sufficient test coverage a mission impossible.

Meanwhile, for the same reason, to best ensure the software quality, developers need identify the most error prone logic for better prioritization between different components.

In this paper, we will introduce the automatic Unit level test framework used in intel media driver development for quarters. It can automatically identify the most critical code block, provide the recommended test coverage for each component, and automatically generate >80% ULT code (~400K Lines of test code) as well as ~35% test cases (~7K test cases) for Intel media driver. It helps us to greatly shrink the average ULT development effort from ~24 Man hours to ~3 Man hours per 1000 Lines of driver source code.

## 2. Auto Unit Level Test Framework

The Auto Unit Level Test Framework for Intel Media driver is composed of the Server-side and the client-side facilities. The server-side of Auto ULT Framework is in charge of automatic error Prone logic detection and the per component test coverage recommendation generation. While the client-side will automatically generate the test code and test cases based on the recommendation from server side. The overall architecture of Auto ULT framework can be shown by Figure1 below.

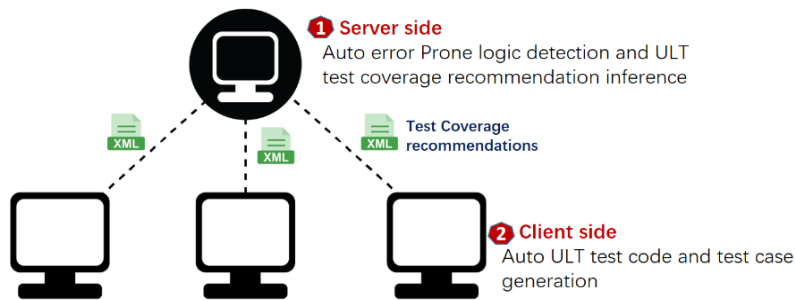

Figure1: Auto Unit Level Test Framework Overview

### 2.1 Auto Error Prone Logic Detection and Test Coverage Recommendation Inference

To better expose the potential issues with limited test cases, different components need be treated differently on their ULT test coverage and those error-prone logics requires better test coverage.

At server side of the Unit Level test framework, it includes one machine learning based inference system which can help to identify the error prone logic based on the recent bug trend and then infer the per component ULT test coverage recommendation.

The basic working flow of the Auto ULT framework Server-side can be shown by figure 2.

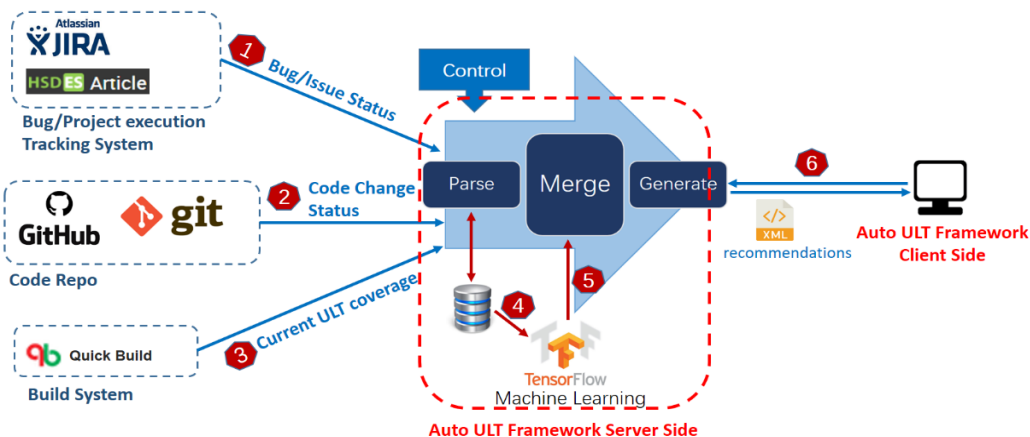

Figure 2: Auto ULT Framework Server-side Working Flow

The detailed steps are as below:

<1> The Automatic ULT framework will regularly (monthly in our case) poll the Bug system for recent bug status. For each new bug, it can get the culprit commit ID from the bug system.

<2> With the culprit commit id listed in the bug, the Automatic ULT framework will query the code repo system for the components / code blocks impacted by the culprit commit.

<3> Meanwhile, from the build the system, the automatic ULT framework can get the current ULT coverage status per each component/code block.

<4> Then by querying the internal database, the Automatic ULT framework can generate the trend for bug V.S. ULT coverage per each component.

<5> At backend of the server-side, we have one pre-trained machine learning system against tensor flow. It can deliver the inference from the trend data got in step 4 and generate the recommendation of the optimal ULT coverage for each component.

<6> The Client side will automatically query for the updated ULT coverage recommendation every time when it launched. If any test coverage improvement required per the new recommendation, it will be highlighted at client side.

The working flow of the machine learning based inference to ULT coverage recommendation mentioned in step 5 can be shown by Figure3 below.

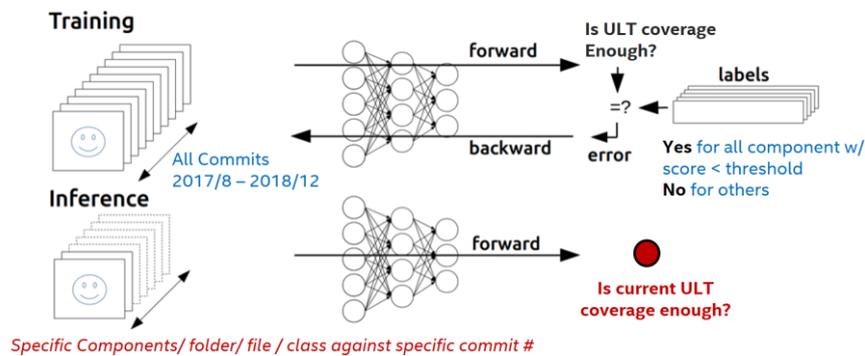

Figure 3: Machine Learning based per component ULT Coverage recommendation

The ULT coverage recommendation can then be used by developers to decide the ULT design and coverage target.

Per the recent survey to all internal media driver developers, >90% interviewees agree upon the accuracy of the recommendation.

## 2.2 Auto Test Generation

After nailing down the ULT design target, developers can then leverage the auto ULT framework on test code and test case generation.

The basic flow for Auto ULT test generator can be shown by Figure4 below.

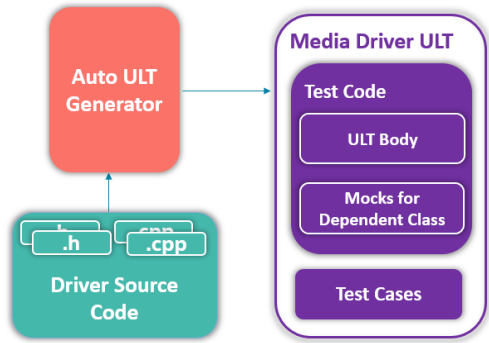

Figure 4: Client side: Auto ULT Test code and Test Case Generator

### 2.2.1 Auto Test Code Generation

Comparing to Driver under Test, ULT is usually of simple logic and fixed pattern which makes the auto ULT code generation feasible.

Let us see what kind of test code is required in a common ULT.

To apply the finer granularity unit level test onto one class, usually it requires several facility classes: one text fixture class, one or more test classes and one or more mock classes.

Figure5 below shows one typical example. Let's say Class A (with one dependent class: class C) is the class under test. To apply the class level ULT onto Class A, we need one test fixture class A_TestCase, one test class Test_A and one mock class to its dependency Mock_C.

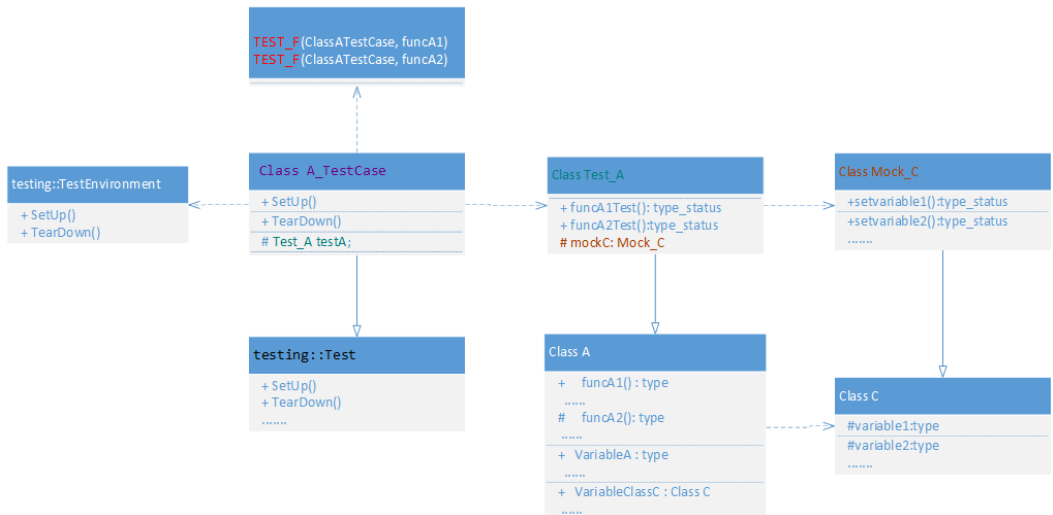

Figure 5: Test Code Required for Finer Granularity ULT

**Class under test** can be as below,

```
class A
{
    ......
    void func1();
    void func2();
    int variable1;
    int variable2;
```

```cpp
       .......
};
```

Then the **test fixture class** *A_TestCase* will look like below

```cpp
class A_TestCase : public testing::Test
{
public:
    virtual void SetUp();
    virtual void TearDown();
    Test_A *testA;
    .......
};

TEST_F (A_TestCase, function1)
{
    testA->function1Test();
}
TEST_F (A_TestCase, function2)
{
    testA->function2Test();
}
```

In the test fixture class, firstly there are Setup() and TearDown() functions including the preparation & cleanup operations for the unit level test.

Secondly it needs include a set of isolated test cases. Each test case will then call into its correspondent in the **test class** *Test_A,* to deliver the real test.

The **test class** *Test_A* is inherited from the **class under test** class A and will provide the real test implementation, including the parameter & logic check. It will be called by the test fixture.

Meanwhile, to achieve the conditional coverage for Class A, we also need a **mock class** for its dependent class *ClassC* which looks like below:

```cpp
class C
{
   ......
   .......
   int variable1;
   int variable2;
   .......
   .......
};
class MOCK_C : public C
{
   .......
   .......
   void SetVariable1();
   void SetVariable2();
   .......
   .......
};
```

From the above description, we can see most of the above code snippets are similar & repetitious. The only exceptions could be the test fixture methods, Setup() and TearDown() which may require some case by case customizations.

Based on above patterns, ULT test code can be easily auto generated from the source code of driver under test.

The high-level blocking diagram for Auto ULT test code Generation can be shown by Figure6 below.

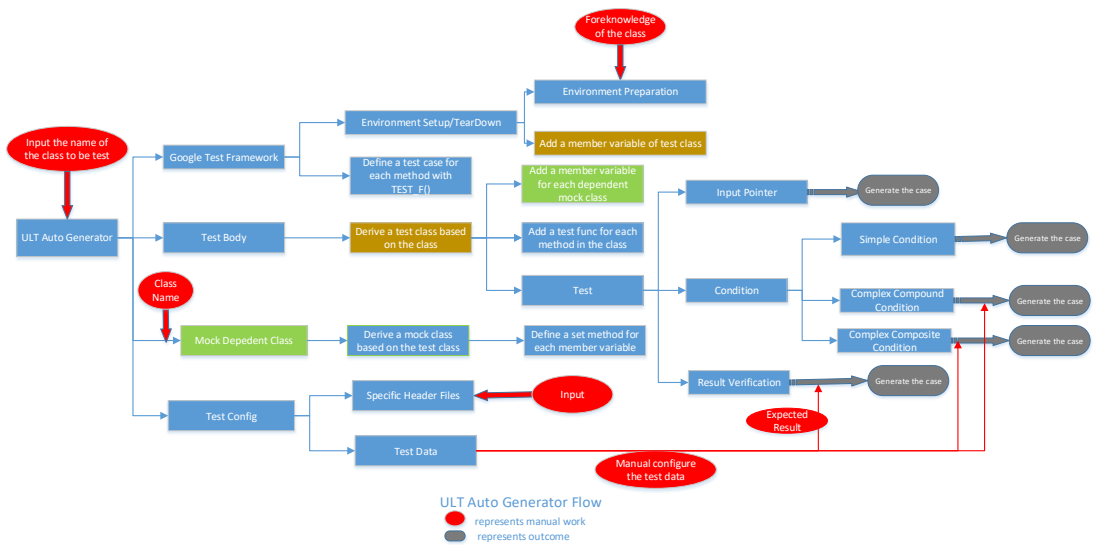

Figure6: High level blocking diagram for Auto ULT test code Generation

Based on above design, now >80% test code (~400K LOC#) for Intel Media driver ULT can be auto generated by Auto ULT Framework.

### 2.2.2 Auto Test Case Generation

Test case design is the most important portion in ULT implementation which we believe deserves more developer involvement. Our strategy on media driver test case design is to have developers focusing on the core logic's test case design for functionality assurance, while try to leverage automation on robustness check for those "else" paths.

In Media driver ULT development, functionality test cases for core logic mainly come from the manual effort. Developers is allowed with the explicit control on interesting members of the input parameters through some pre-defined configuration files, and the Auto ULT framework will help on corresponding initialization code generation in test class/mock class based on the settings in configuration files.

Robustness assurance is another important goal for our ULT. Unlike functionality, robustness is more decided by the code quality of so-called "else" paths, and effective robustness check usually requires higher ULT coverage till condition/decision level. Also, usually robustness check just needs ensure no crash or assert been triggered during the test and does not require any reference. All above specialties make robustness test cases more suitable to be auto generated.

In Auto ULT framework, to promise the sufficient condition/decision coverage, Auto ULT Framework will try to add robustness test cases based on the following policies:

<1> For those conditions decided by input parameters, auto ULT framework will leverage the fuzzing to generate the input parameters for corresponding test cases.

<2> For those conditions decided by return value of the nested function call, auto ULT framework will generate the required fake return inside the mock classes.

For Robustness check, currently we use the condition/decision coverage as the metrics to measure the quality for auto-generated test cases. We believe as long as we got sufficient coverage on conditions/decisions, most of the robustness issues should be well exposed.

Now in total we can have ~35% test cases (~7K cases) auto generated by the auto ULT framework requiring almost no extra changes from developers.

## Summary

Unit level test is an important approach to reduce the defect and improve the software quality. For Intel Media driver, we have the ULT coverage target of 100% functional coverage and >70% conditional coverage. With the help from ULT, in 2018, Intel Media driver achieved 0 Quality Events and OEM driver escapes is dramatically reduced by 36%.

But manual ULT development can lead to big extra development effort. To better offload developer from the tedious ULT development task and focus on the real test case design/verification, we designed the Auto ULT framework. It can help our developer easily identify the coverage gap based on bug trend and automatically generate >80% ULT code (~400K Lines of test code) as well as ~35% test cases (~7K test cases).

With the help of Auto ULT framework, our average ULT development effort has been greatly shrunk from ~24 Man hours to ~3 Man hours per 1000 lines of driver source code.

## Acknowledgements

Thanks to all colleagues working on refactoring for continuous software delivery and competency improvement. Appreciate your hard work to turn all our good designs into the reality.

**Authors**

Ning Luo is the senior software architect at Intel. His research interests include software requirements and architecture, continuous delivery, DevOps, and software product lines. Please contact him at ning.luo@intel.com.

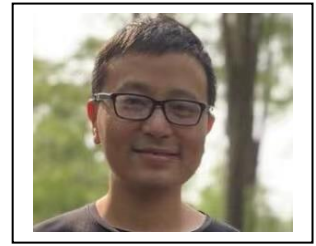

Linlin Zhang is a senior software engineer at Intel. Please contact her at livia.zhang@intel.com

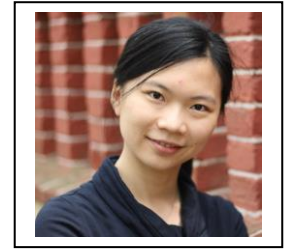